\documentclass[prd,a4paper,superscriptaddress,nofootinbib]{revtex4}
\usepackage{amsmath,epsfig,natbib,bm}

\begin{document}


\newcommand{\Deltathree}{\Delta}
\newcommand{\DeltaX}{X_s}
\newcommand{\Ohat}{\hat{O}_n}
\renewcommand{\chi}{{\cal X}}
\newcommand{\N}{\chi}
\newcommand\K{{\cal K}}
\newcommand{\dphinv}{\delta\phi_{\text{gi}}}
\newcommand{\adph}{{\tilde{Q}}}
\newcommand{\aPiphi}{\Pi_\adph}

\newcommand\be{\begin{equation}}
\newcommand\ee{\end{equation}}

\renewcommand{\nu}{n}

\renewcommand\({\left(}
\renewcommand\){\right)}
\renewcommand\[{\left[}
\renewcommand\]{\right]}
\newcommand\lagrange{{\cal L}}
\newcommand\cll{{\cal L}}
\newcommand\del{\nabla}
\newcommand\Tr{{\rm Tr}}
\newcommand\half{{\frac{1}{2}}}

\newcommand\calS{{\cal S}}
\renewcommand\H{{\cal H}}

\renewcommand\P{{\cal P}}
\newcommand{\la}{\langle}
\newcommand{\ra}{\rangle}
\newcommand{\phpr} {{\phi'}}
\newcommand{\gam}{\gamma_{ij}}
\newcommand{\sqgam}{\sqrt{\gamma}}
\newcommand{\delk}{\Delta+3{\cal K}}
\newcommand{\dph}{{\delta\phi}}
\newcommand{\phidot}{{\dot{\phi}}}
\newcommand{\om} {\Omega}
\newcommand{\dom}{\delta^{(3)}\left(\Omega\right)}
\newcommand{\rar}{\rightarrow}
\newcommand{\lrar}{\leftrightarrow}
\newcommand{\Rar}{\Rightarrow}
\newcommand{\labeq}[1] {\label{eq:#1}}
\newcommand{\eqn}[1] {(\ref{eq:#1})}
\newcommand{\labfig}[1] {\label{fig:#1}}
\newcommand{\fig}[1] {\ref{fig:#1}}
\newcommand{\Omtot}{\Omega_{\mathrm{tot}}}
\newcommand{\Omk}{\Omega_{\mathrm{K}}}

\title{Closed Universes from Cosmological Instantons}

\author{Steven Gratton}
 \email{sgratton@princeton.edu}
 \affiliation{Joseph Henry Laboratories, Princeton University,
 Princeton, NJ 08544, USA} 
\author{Antony Lewis}
 \email{Antony@AntonyLewis.com}
 \affiliation{DAMTP, CMS, Wilberforce Road, Cambridge CB3 0WA, UK}
\author{Neil Turok}
 \email{N.G.Turok@damtp.cam.ac.uk}
 \affiliation{DAMTP, CMS, Wilberforce Road, Cambridge CB3 0WA, UK}

\begin{abstract} 
\vspace{\baselineskip}

Current observational data is consistent with the universe being
slightly closed.  We investigate families of singular
and non-singular closed
instantons that could describe the beginning of a closed inflationary universe.
We calculate the scalar and tensor perturbations that would
be generated from singular instantons and compute the corresponding
CMB power spectrum in a universe with 
cosmological parameters like our own. We investigate 
spatially homogeneous modes of the instantons,
finding unstable modes which render the instantons sub-dominant
contributions in the path integral.  We show that a suitable
condition may be imposed on singular closed instantons, constraining
their instabilities. With this constraint these instantons
can provide a suitable model of the early universe, and predict CMB
power spectra in close agreement with the predictions of slow-roll inflation.

\end{abstract}

\maketitle

\vskip .2in

\section{Introduction}

Recent measurements of the Cosmic Microwave Background (CMB)
temperature anisotropy~\cite{Bernardis01,Stompor01} have been able to
tightly constrain the curvature of the universe. The results indicate that the
universe is close to being spatially flat, with current limits of
$0.94<\Omega_{\mathrm{tot}}<1.06$ at one sigma. However closed models
still take up a significant part of the possible parameter space, and
unless the universe is significantly open we shall probably never be
able to rule out closed models. In this paper we discuss an 
approach to quantum cosmology which yields closed inflationary
universes consistent with
the current observations.

Over the past three years there has been a development of methods
treating open
universes~\cite{Garriga98,HT,Gratton99,Hertog00,Gratton00,Gratton00;neg}
in the context of Euclidean quantum cosmology~\cite{Hartle83}.  In this
paper we adapt these methods to closed universes, and use them to
calculate the perturbation power spectrum for such models. The reader
is referred to the above papers for general details; here we mainly
emphasize the new considerations arising from the closed geometry. In
many ways the closed calculations are easier, with the spatial
hypersurfaces being compact and the analytic continuation being
simpler.  A major difference is that a different variable must be used
to represent the scalar fluctuations about the instanton in the closed
case, due to the different ranges of the background variables.

The structure of this paper is as follows.  We firstly describe
non-singular and singular closed instantons and their analytic
continuations to possible universes.  We then introduce variables
suitable for the description of scalar fluctuations across the entire
instantons and show that in both cases negative modes exist which
render these instantons unstable.  For the non-singular case we show
that at least two negative modes exist, excluding these from even
speculative tunneling applications~\cite{Coleman88}.  In the singular
case we are able to motivate a constraint at the
singularities~\cite{Kirklin00}, which projects out these negative
modes.  We then proceed to calculate the predicted scalar and tensor
perturbations about such constrained singular closed instantons.  We
calculate the two-point correlation functions in the Euclidean region
and analytically continue these expressions into the Lorentzian
region.  We compute the power spectra at the end of inflation and use
these as initial conditions for the radiation dominated era in to
order to compute the predicted CMB power spectra for a given
cosmological model. The suitability of these models as descriptions of
the early history of our universe is then commented upon.

\section{Instantons for Closed Universes}
\label{sec-closedinst}
Consider a closed universe with line element
\begin{eqnarray}
ds^2=-dt^2+a^2(t) d\Omega_3^2 
\end{eqnarray}
and scalar field $\phi$ with potential $V(\phi)$.  
The Friedmann Equation reads
\begin{eqnarray}
\(\frac{\dot{a}}{a}\)^2 =\frac{\kappa}{3}\(\frac{1}{2} \phidot^2+
V(\phi)\)-\frac{\K}{a^2} 
\end{eqnarray} 
where $\K$ is the spatial curvature constant, with $\K>0$
corresponding to a closed universe. One sees that it is possible for both $a$ and $\phi$ to be stationary
at the same time, $t=0$ say, if at this moment $a$ and $\phi$ are such
that $\kappa a^2 V /3 =\K$.   Around $t=0$, $a$ and $\phi$ will then
have power series expansions even in $t$.  Hence one may consider the
analytic continuation $\sigma=it$, under which both $a$ and $\phi$
remain real.  The line element then takes the Euclidean form
\begin{equation}
ds^2= d\sigma^2 +b^2(\sigma) d \Omega_3^2 
\labeq{emetric}
\end{equation}
with $b(\sigma)$ the radius of the three-sphere, related to $a$ by 
$b(\sigma)=a(-i\sigma)$. The Einstein and 
scalar field 
equations take the form 
\begin{equation}
\phi_{,\sigma\sigma}+3{\frac{b_{,\sigma}}{b}}\phi_{,\sigma} 
=V_{,\phi},\qquad b_{,\sigma \sigma} = -{\frac{\kappa}{3}} b \(
\phi_{,\sigma}^2 +V\)
\labeq{EEs}
\end{equation}
where $\kappa \equiv 8 \pi G$, 
$\phi_{,\sigma} \equiv \partial_\sigma \phi$ and $V_{,\phi}\equiv
\frac{dV}{d\phi}$.  There are solutions to these equations with
various geometries. For the
open case of  Ref.~\cite{Gratton99} the background metric and scalar
field behaved appropriately at a pole. Here we are interested in
solutions symmetric about an equator (where $b_{,\sigma}=0$) that have continuations
into a real closed universe.

One can introduce conformal coordinates in both the Euclidean
and Lorentzian regions, defined by integrating $dX=-b d\sigma$ and
$d\tau=a dt$ from the matching surface.  In this closed case, since the
scale factor is finite at the matching point, the conformal
coordinates $X$ and $\tau$ behave just like $\sigma$ and $t$ and
can be continued into each other similarly.  This is another
significant simplification when compared to the open case.  

By virtue of the boundary conditions required by the continuation we
see that $b$ and $\phi$ must be symmetrical about the equator of the
instanton.  
For potentials $V$ greater than or equal to zero, evolving the equation
for $b$ from the equator enables one to see that $b$ must twice
reach zero a finite 
distance in $\sigma$ away, at $\pm \sigma_m$ say.  These are poles of
the instanton.  Because of the the symmetry about the equator they
must be of the same form and can either be regular or singular, as
discussed in detail in Refs.~\cite{HT,Gratton99}.   Singular instantons exist
for almost any scalar field potential, whereas the regular instantons
can only exist for potentials which have a local maximum and are
sufficiently negatively curved 
in the region of the peak (from the work of~\cite{Jensen84}
we require that $-3 V_{,\phi\phi}/\kappa V > 10$).  See the illustration in
Fig.~\fig{contil}.

\begin{figure}[t!]
\begin{center}
\psfig{figure=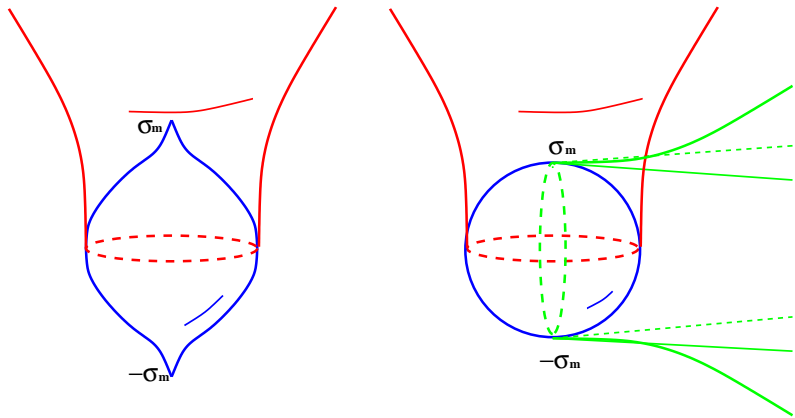,angle=-90,width=10.5cm}
\caption{This figure illustrates the continuation from singular (left) and
non-singular (right) instantons into Lorentzian universes. Analytic
continuation at the equator yields a closed universe. In the
non-singular instanton continuation at the poles gives open universes.\labfig{contil}}
\end{center}
\end{figure}


To reiterate, the key difference between open and closed instantons is
that the former have their boundary conditions imposed by behavior at
a pole whereas the latter have their boundary conditions imposed by
behavior at the equator.  One might ask if these conditions ever
overlap.  Indeed they do, as it is possible to continue any regular
closed instanton not from its equator into a closed universe but from
one of its poles into an open universe.  From the open point of view,
these correspond to the ``multibounce'' instantons mentioned in Sec.~3
of~\cite{Gratton99}.  From the negative mode argument presented there, one
immediately suspects that these solutions will have more than one
negative mode and hence not be relevant for cosmological applications.
Indeed, we will prove below that a large class of these non-singular
models have at least two negative modes.  
   
We now turn to investigate the scalar fluctuations about these
instantons.  

\section{The Second Order Action}

Our starting point is the second order
action for scalar perturbations in the Lorentzian universe, as
discussed in Sec. 4 of~\cite{Gratton99}, to which we refer the reader for
definitions.  
The result we use agrees with that given in~\cite{Mukhanov92} up to a
surface term, and is reproduced here, 
\begin{eqnarray} S_2 & = & \frac{1}{2\kappa} \int d\tau d^3 x a^2 \sqrt{\gamma} \bigg\{
  -6\psi'{}^2-12{\cal H} A\psi'+2\Delta\psi\(2A-\psi\)-2\({\cal
    H}'+2{\cal H}^2\)A^2
 \nonumber \\
& & +\kappa\(\dph'{}^2+\dph\Delta\dph-a^2 V_{,\phi\phi}\dph^2\)
+2\kappa\(3\phpr\psi'\dph-\phpr\dph'A-a^2V_{,\phi}A\dph\) \nonumber \\
& & +{\cal K}\(-6\psi^2+2A^2+12\psi A + 2\(B-E'\)\Delta\(B-E'\)\) \nonumber \\
& &  +4\Delta\(B-E'\) (\frac{\kappa}{2} \phpr\dph-\psi'-{\cal
    H}A) \bigg\}.  \labeq{2acnew} \end{eqnarray}
  This is well defined for all values of $\phpr$ and the three-space
  Laplacian $\Delta$.  In a closed universe $\K > 0$ and 
the eigenvalues of $\Delta $ are given by $-\K(n^2-1)$, where $n$ is a
positive integer.

For the spatially inhomogeneous modes $(n\geq 2)$ we introduce the momenta
canonically conjugate to $\psi$, $E$, and $\dph$ and rewrite the
action in first-order form, as in Eq.~(20) of~\cite{Gratton99}.  Then we
integrate over $B$, leaving us with a delta functional for $\Pi_E$.
Integrating over this effectively sets $\Pi_E$ to zero.  
For the homogeneous $(n=1)$ mode $B$ and
$E$ disappear from the action, so we can only introduce momenta
canonically conjugate to $\psi$ and $\dph$.  However the action can still be
rewritten in first-order form, and has the same form as in 
the inhomogeneous case.  So we now have the following expression for
all modes in
terms of $\psi$, $\Pi_\psi$, $\dph$, $\Pi_{\dph}$, and $A$:
\begin{eqnarray}
S_2 &=& \int d\tau d^3 x \Bigg\{\Pi_\psi \psi'+ \Pi_{\dph} \dph'
+\frac{\kappa\K} {4 a^2 \sqgam \(\delk\)}\Pi_{\psi}^2-\frac{1}{2 a^2
\sqgam}  \Pi_{\dph}^2 - \frac{\kappa}{2}\phpr\Pi_{\psi} \dph \nonumber \\
& &~~~ - \frac{a^2\sqrt{\gamma}}{\kappa} \(\(\Delta+3{\cal K}\)\psi^2
  -\frac{\kappa}{2}\(\Delta+3{\cal K}
  -\H^2-\H'+\frac{\phi'''}{\phpr}\)\dph^2\) \nonumber \\
& & ~~~- A
\(-\H\Pi_\psi+\phi'\Pi_{\dph}+\frac{2a^2\sqgam}{\kappa}\(-\(\Delta+3{\cal
K}\)\psi+\frac{\kappa}{2}\(\H\phpr-\phi''\)\dph\)\)\Bigg\}.
\labeq{oneact}
\end{eqnarray}
 The $n=2$ mode corresponds to a gauge degree of freedom and is ignored as
discussed in Sec.~4 of~\cite{Gratton00;neg}.  
We now have some choice in deciding which linear combination of
variables to use in order to describe the single scalar
degree of freedom of the system.  As we shall see, different
choices are particularly suited to different background solutions.   

\section{Non-singular Closed Instantons}
 
From our experience~\cite{Gratton00;neg} with Hawking-Moss~\cite{Hawking82}
and Coleman-De Luccia~\cite{Coleman80}  
instantons
one expects a variable related to $\dph$ to be a good variable to use
in treating closed instantons.  We remind the reader that, as discussed
in detail in~\cite{Gratton00;neg}, a good variable is one whose Euclidean
action is bounded below for all normalized fluctuations of that
variable about the instanton in question.  This corresponds to having
a positive kinetic term in the Euclidean action for all
values of $\Delta$.  

We discussed closed instantons in Sec.~\ref{sec-closedinst}.  The
simplest instanton is that of Hawking and Moss, which has the 
scalar field constant and at a local maximum of the potential.
It is a round $S^4$ and  has $O(5)$
invariance.  It can be analytically continued to either the closed or
the open slicing of De Sitter space.  The closed instantons
discussed in Sec.~\ref{sec-closedinst} may, as discussed 
above, also be continued to open or closed inflationary universes.

Under an open continuation, we
have~\cite{Gratton99}
\begin{eqnarray}
\sigma \lrar i t,~~~ X\lrar -\tau -i \pi /2, ~~~ b^2 \lrar
-a^2,~~~\Delta \lrar -\Delta,~~~\K \lrar \K,
\end{eqnarray} 
whereas under a closed continuation,
\begin{eqnarray}
\sigma \lrar i t,~~~ X\lrar i\tau, ~~~ b^2 \lrar
a^2,~~~\Delta \lrar \Delta,~~~\K \lrar \K,
\end{eqnarray}
as discussed above.  The differing analytic
continuations ensure that if one starts in the Lorentzian region in
either an open or a closed slicing, one gets an identical expression when
continued back to the Euclidean region.  For example, the combination
$\Delta +3 \K$ occurs in both Lorentzian regions.  This continues
to $-\Delta+3 \K$ in the Euclidean region from an open universe, or to
$\Delta+3\K$ from a 
closed universe.  However, $\K$ is negative in an open universe, so
these two expressions really differ only by a sign.  In fact,
the sign is also mopped up by another minus sign originating 
from $a^2$ so that both continuations produce an  identical Euclidean
action,
 as long as
we interpret $\K$ in the Euclidean region to really be $| \K |$ of
the Lorentzian region.  Since this paper only deals with closed
universes, for which $\K$ is positive, 
we can ignore this subtlety.

Sec.~4 of~\cite{Gratton00;neg} introduced a variable which made the
negative mode structure transparent. Here we attempt to adapt the
argument to the case of a closed instanton.
We introduce the gauge invariant perturbation variable $\dphinv$,
\begin{eqnarray}
\dphinv= \dph-\frac{\kappa \phpr\Pi_\psi}{2 a^2 \sqgam(\delk)}.
\end{eqnarray}
After integrating out the other variables from~\eqn{oneact},
analytically continuing 
to the Euclidean region, and introducing the rescaled variable $Q=b
\dphinv / \phi'$ we obtain the action
\begin{eqnarray}
S_2=\half\int dX d^3x \sqgam \phi'{}^2
\(\frac{Q'{}^2}{1-\frac{\kappa
    \phi'{}^2}{2\(\Deltathree +3\K\)}}-\(\Deltathree +3\K\)Q^2\).
\labeq{cdelact} 
\end{eqnarray}
One might think that if $1-\kappa \phi'{}^2 /6\K >0$ across the
instanton, so that the $Q'{}^2$ always has a positive coefficient,
$Q=\mathrm{const}$ might be a homogeneous negative mode.  However,
there are two subtleties in the closed case which did not occur in the
parallel case for open-only instantons (for which $\phi'$ is always
positive.)  Firstly, because the scale factor is finite on the matching
surface and $\phi'$ is zero there, $Q$ is not well defined in the
closed case. Secondly, the natural measure in the
open-only case was $\phi'{}^2$, but in the closed case this vanishes on the
equator and so is inadmissible.
However the variable $\adph \equiv b\dphinv$ is well defined, and we can
proceed to derive the action in terms of this variable.

Starting from Eq.~\eqn{oneact} we perform the path integral over $A$
which imposes a delta functional allowing us to express $\psi$ in terms of $\Pi_\psi$ and
$\Pi_\dph$. After integrating by parts to remove terms in the
derivatives of the conjugate momenta we define $\aPiphi$ as the variable
conjugate to $\adph$. Substituting $\Pi_\dph$ for $\aPiphi$ we find the
action is independent of $\Pi_\psi$, and so its path integral can be
neglected as an infinite gauge-orbit volume. Finally performing the Gaussian
integral over $\aPiphi$ we obtain the Euclidean action
\begin{eqnarray}
S_2 &=& \half\int \frac{dX d^3 x
\sqgam}{1-\frac{\kappa\phi'{}^2}{2(\Delta+3\K)}}\left( \adph'{}^2 - \left[
\Delta+3\K - \half\kappa\phi'{}^2 -\frac{\phi'''}{\phi'} -
\frac{\kappa(\phi'')^2}{\Delta+3\K - \half\kappa\phi'{}^2}\right]\adph^2\right).
\labeq{closedact}
\end{eqnarray}
%
Note that $\phi'''/\phi'$ is well
behaved at the equator because $\phi$ is an even function for the
closed instanton, and this action is suitable as long as $1-\kappa
\phi'{}^2 /6\K >0$ across the entire 
instanton.  One can confirm this numerically in any particular case of
interest---most regular closed instantons leading to reasonable amounts
of inflation satisfy this condition. 

As expected one can confirm that $\adph=\phi'$ is a spatially homogeneous negative mode of the
fluctuation operator associated with this action.
However, $\phi'$ is odd about the equator so this negative
mode has a node.  The eigenvector associated with
the lowest eigenvalue must be nodeless and hence there must exist
a symmetric mode function with a more negative eigenvalue.
(This can be confirmed numerically by substituting a symmetric test
function, $\adph=b$ say, into~\eqn{closedact} and checking that the
action integral over the instanton is negative.)  So
we have proved that these instantons must have at least two negative
modes, and therefore cannot have anything to do with
tunneling~\cite{Coleman88,Gratton00;neg}.  This
confirms explicitly the arguments based 
on spectral flow from the Hawking-Moss instanton given in Sec.~3
of~\cite{Gratton00;neg}.   

As far as we are aware, there is no way to constrain these instantons to
eliminate their instabilities. Thus there is no motivation for considering
them as being significant in the Euclidean quantum path integral.  No
tunneling role being apparent either, we do not consider these
instantons further.  Instead we now turn to singular instantons, where
suitable constraints may be applied.

\section{Singular Closed Instantons}

On singular instantons, as the scalar field runs away to infinity at
the poles, the condition $1-\kappa\phi'{}^2 / 6\K>0$ is certainly
violated.  This means that $\dphinv$ is not a suitable variable to use
when investigating fluctuations around closed singular instantons.

%


A good physical choice of variable is the gauge invariant combination
\begin{equation}
\chi\equiv \psi+\frac{\H}{\phi'  } \dph,
\end{equation}
the comoving curvature perturbation. In the gauge
$\dph=0$ this is just $\psi$. Notice that in the definition the factors
$\H$ and $\phi'$ are combined in such a way as to be finite at the
equator of the instanton, since in this closed case both go linearly
to zero there.

Using the definition we substitute $\dph$ for $\chi$ in
Equation~\eqn{oneact}.  The path integral over $A$ imposes a delta functional which allows us to express $\psi$ in terms of $\Pi_\psi$ and
$\Pi_\dph$. After integrating by parts to remove terms in the
derivatives of the conjugate momenta we define $\Pi_\chi$ as the variable
conjugate to $\chi$. Substituting $\Pi_\dph$ for $\Pi_\chi$ we find the
action is independent of $\Pi_\psi$, and so its path integral can be
neglected as an infinite gauge-orbit volume. Finally performing the Gaussian
integral over $\Pi_\chi$
%
%
and continuing to the Euclidean region we are left with
\begin{eqnarray}
S_2 = \frac{1}{\kappa} \int dX d^3 x \frac{b^2
\sqgam z^2}{1-z^2\K/(\Deltathree +3\K)} \( \N'{}^2
-\left[\frac{2\K}{1-z^2\K/(\Deltathree +3\K)}\frac{z'}{z \H}+\Deltathree  +4\K \right]
\N^2\)
\labeq{nact}
\end{eqnarray} 
where we have introduced $z \equiv \sqrt{\kappa/2} \phi'/\H$ for
clarity.

Let us now examine the behavior of this action across the
instanton.  For $\N$ to be a suitable variable we need the coefficient
of the kinetic term to be positive everywhere.  Remembering that the
eigenvalues of $\Deltathree $ are $-\K(n^2-1)$, where $n$ is a
positive integer (and $\K$ is a positive real number in this closed
case), we see that 
this is indeed the case for spatially
inhomogeneous fluctuations. In the homogeneous case ($n=1$) it will be
positive if $1-z^2/3 >0$ across the instanton.  This
can be checked numerically for any case of interest, but we can also
make analytic arguments for wide classes of cosmological
instantons.  Let us first examine behavior near the equator.  From
the scalar field equation in proper Euclidean time, we have $\phidot
\approx V_{,\phi_0} \sigma$ and from the equation for the scale factor
we have
$\dot{b} \approx -\kappa b V(\phi_0) \sigma /3$.  Hence $1-z^2/3
\approx 1- 3 V_{,\phi_0}^2/2 \kappa V(\phi_0)^2$.  For a monomial
potential $\lambda \phi^r$, we can link the starting value $\phi_0$ of
the field with the number $P$ of e-foldings by the
relation~\cite{Bucher95} $\kappa \phi_0^2 =2 r P$, leading to the
requirement that $P>3r/4$.  For cosmologically interesting solutions,
we require $P \sim 50$, and so for $r\sim 1$ the kinetic term is
positive.  Now let us
examine the behavior near the singularities.  Using the
background field equations we have the relation
\begin{eqnarray}
1-\frac{z^2}{3} = \frac{1}{\H^2} \( \K - \frac{\kappa b^2 V}{3}\).
\end{eqnarray}
If $\DeltaX $ is the conformal distance from a singularity, $b^2$
goes like $\DeltaX $ and $\phi$ goes like $-\sqrt{3/2\kappa} \ln
\DeltaX $.  So $\H^2$ goes like $1/4 \DeltaX ^2$ and we see that
$1-z^2/3$ simply goes like $4\K \DeltaX ^2$, staying positive.  
See Fig.~\fig{ant1} for a graph showing the behavior of $b$ and $z^2$
across a typical instanton.

\begin{figure}[t!]
\begin{center}
\psfig{figure=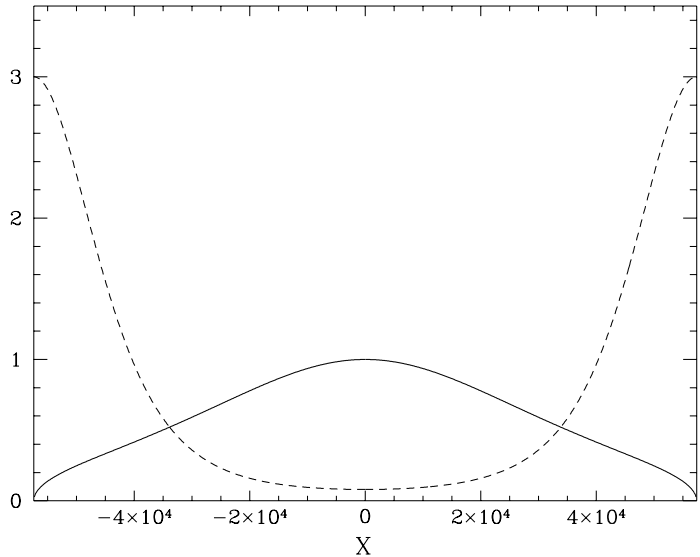,angle=-90,width=10.5cm}
\caption{The scale factor (solid line) and $z^2$ (dashed line)
versus Euclidean conformal time for a
closed singular instanton with an $\half m^2\phi^2$ potential. The
closed universe is obtained by analytic continuation on the
central hypersurface, and the singularities are at the two end points.\labfig{ant1}}
\end{center}
\end{figure}

It remains to examine the behavior of the
potential term across the instanton. At the equator
$z'/z\H=\dot{z^2}/2 z^2 H$ tends to
a finite constant, since $z^2$ is even in $\sigma$ and has an
$O(\sigma^2)$ term.  Near a singularity the same term vanishes like
$-8\K \DeltaX ^2$.  Hence the potential tends to a finite constant, and
furthermore in the special case of the spatially homogeneous mode it tends
to zero.  The curvature perturbation $\N$ is therefore a suitable variable to use for investigating a wide range of
singular closed instantons.  

We must now address the question of instabilities of these instantons,
which we do by examining the behavior of mode functions near the
singularity.  

\section{Stability of Closed Singular Instantons} 

The idea of Euclidean quantum cosmology is that the Euclidean path
 integral should uniquely determine the quantum state of the
universe. If the instantons discussed above are to be a useful
approximation, it must be that the Euclidean action for
small fluctuations about these instantons uniquely determines
the set of allowed fluctuation modes. 

Consider first the inhomogeneous modes.  The mode equation associated
with the action~\eqn{nact} above takes the form $(\DeltaX \N')'
\approx 0$ near a singularity, with general solutions of the form $A
\ln \DeltaX +B$.  Substituting back into the expression for the
action we see that the logarithmically divergent solution has positive
infinite Euclidean action and hence is totally suppressed by the
Euclidean path integral.  Hence the Euclidean action uniquely
determines the state of the spatially inhomogeneous modes.

For the homogeneous modes the equation takes the
form $ (\N' / \DeltaX)' \approx 0$, with general solutions of the form
$A \DeltaX^2 +B$.  Now however, substituting back we find that both
solutions have finite action.  Hence the Euclidean path integral cannot
uniquely determine the state of the homogeneous fluctuations.  This
tells us that singular closed instantons are not indeed true saddle
points of the Euclidean action, as might have been expected from
experience with the singular open instantons of~\cite{Gratton00;neg}.  However,
as there, one can proceed to introduce a constraint at each of the
singularities.  A suitable constraint is provided by the work of
Kirklin, Turok and Wiseman~\cite{Kirklin00}, as applied in~\cite{Gratton00;neg}.
This corresponds to fixing $m\equiv b e^{\sqrt{\kappa/6} \phi}$ at each
singularity.  We must have the same value of $m$ at each singularity
in order that the instanton be symmetric about the equator.
Each value of $m$ effectively corresponds to a given value of $\phi$
at the equator via the background field equations.  So we are left
with a one-parameter family of instantons, corresponding to Lorentzian
universes with different starting
points for the scalar field in its potential.  We are free to set this
parameter to give a universe like that which we see today.   

We then require that fluctuations satisfy $\delta m=0$ at
the singularities.  In the gauge $\dph=0$ this
condition reduces to $\delta b /b =0$ at the singularities.  But this,
up to a constant, is just our $\N$.  Hence we require that the
homogeneous modes behave like $\DeltaX^2$ near the singularities.  We
can now numerically solve this eigenproblem of finding the lowest
eigenvalue such that the associated eigenmode satisfies these boundary
conditions.  It has turned out that for the instantons we have
considered that this eigenvalue has indeed been positive.  Hence the
constraint has projected out the unstable negative modes.  

Thus closed singular instantons, so constrained, have been shown to
provide a sensible starting point for a perturbation expansion
and we now proceed to calculate
the correlators for fluctuations about these instantons,
and their observational consequences for the CMB.

\section{The Scalar Power Spectrum}


To work out the observable predictions from the instanton we wish to
compute the power spectrum for the perturbations at the end of
inflation, and relate it to the power spectrum in the early radiation
dominated era. We define the power spectrum so that the gradient of the 3-Ricci
scalar on co-moving hypersurfaces (i.e. in the rest frame of the total
energy, denoted by a tilde) receives power
\begin{eqnarray}
\la | \tilde{D}_a \tilde{R}^{(3)}|^2\ra = \sum_{n} \frac{n}{n^2-1}
\frac{16 k^6}{a^6} \left(\frac{n^2-4}{n^2-1}\right)^2 \P_\chi (n),
\end{eqnarray}
where the $k^2=\K(\nu^2-1)$. At the end of inflation the curvature perturbation variable $\chi$ is
simply related to the perturbation in the 3-Ricci scalar, giving
\begin{eqnarray}
\la |\chi|^2\ra =   \sum_n \frac{n}{n^2-1} \P_\chi (n).
\end{eqnarray}
This convention for the power spectrum is the natural closed analogue
of that given in~\cite{Lyth95}, where in the flat space limit a scale
invariant spectrum corresponds to $\P_\chi = \text{constant}$.
Since we are only considering single field inflation the perturbations
will be adiabatic and the curvature perturbation is conserved on
super-Hubble scales. The power spectrum is therefore time independent,
and what we compute at the end of inflation will accurately predict
the 
super-Hubble power spectrum in the radiation dominated era.

It is convenient to do a harmonic expansion in terms of normalized
eigenfunctions $Q_{nlm}$, where 
$$
\Deltathree Q_{nlm} = -\K(n^2-1)Q_{nlm} \nonumber
$$
\begin{eqnarray}
\int d^3 x \sqrt{\gamma}\, Q_{nlm}(x) Q_{n'l'm'}(x) = \delta_{nn'}\delta_{ll'}\delta_{mm'}.
\end{eqnarray}
The curvature perturbation is then expanded in terms of its modes as
\begin{eqnarray}
\chi(x,\tau) = \sum_{nlm} \chi_{nlm}(\tau) Q_{nlm}(x).
\end{eqnarray}
Inserting the mode expansion into the second
order action and integrating by parts the action takes the form
\begin{eqnarray}
S_2 = \sum_{nlm} \half\int dX \,\chi_{nlm} \Ohat \chi_{nlm},
\end{eqnarray}
where $\Ohat$ is a second order linear differential operator.
 The action for all modes with the same
eigenvalue is equivalent, so we drop the $l$ and $m$ subscripts. The
power spectrum that we wish to compute will be given 
by the correlator for the modes $\chi_n$. Doing the sum over the
degenerate modes of the same eigenvalue we obtain
\begin{eqnarray}
\P_\chi(n) = \frac{\K^{3/2}n(n^2-1)}{2\pi^2} \la | \chi_n|^2\ra.
\end{eqnarray}
For perturbations we are only interested in $\nu\ge 3$, and the factor
of $2\pi^2/\K^{3/2}$ is just the volume of the three-sphere.

We compute the Lorentzian Green function $G_n(\tau,\tau)=\la | \chi_n(\tau)|^2\ra$ by analytic continuation from the
Euclidean Green function.
To compute the correlator we need to solve the equation for the Euclidean Green function 
\begin{eqnarray}
\Ohat G_{E,n}(X,Y) = \delta(X-Y).
\end{eqnarray}
The operator $\Ohat$ is in Sturm-Louville form $\Ohat G =-(pG')' - pqG$,
and we can therefore construct the Green function from two classical
solutions. To give the delta function at $X=Y$ the classical solutions
are joined together with a change in derivative of $-1/p$.  As discussed in the
section above, the Euclidean action infinitely suppresses the
logarithmically divergent solution near a singularity, and so $G_{E,n}$ is
constructed from two classical solutions $\N_-$ and $\N_+$  satisfying
Neumann boundary conditions at $X=-X_m$ and $X=+X_m$ respectively.
For $X\le Y$we have
\begin{eqnarray}
G_{E,n}(X,Y)= \frac{\N_+(Y) \N_- (X)}{\(p(\N'_- \N_+ - \N_- \N'_+)
\)|_{Y}}
\labeq{gn}
\end{eqnarray}
with the corresponding result for $X\ge Y$.  From the
differential equation the denominator is in fact independent of $Y$
and is conveniently evaluated on the equator at $X=0$.  By symmetry,
we may choose $\N_- (X)=\N_+ (-X)$, and so the
denominator just becomes $2 p \N_- \N'_-$ evaluated at $X=0$.  
We now have an
expression for the Euclidean correlator which may be analytically continued into the closed
Lorentzian universe. 

\section{Analytic Continuation}

To continue the expression for the correlator to the Lorentzian universe, we need to set
$X=i\tau$.  For small $X$ and $\tau$ we have
\begin{eqnarray}
\N_- (X) \approx \N_- (0) + \N'_- (0) X  \rar \N_-^L (\tau) \approx \N_- (0) + i \N'_- (0) \tau
\end{eqnarray}
and $\N_-^L$ is therefore complex. At later times it will have the form
\begin{eqnarray}
\N_-^L (\tau) = \N_e(\tau) + i \N_o(\tau)
\end{eqnarray}
where the real part is even [$\N_e'(0)=0$, $\N_e(0)=\N_- (0)$], and the
imaginary part is odd [$\N_o(0)=0$, $\N'_o(0) = \N'_-(0)$]. 
Dividing through by the denominator $2 p \N_-
\N'_-$ and taking the equal time limit of the Lorentzian Green function
we get
\begin{eqnarray}
G_n(\tau,\tau) =\la | \chi_n(\tau)|^2\ra=\frac{\N_e(\tau)^2 + \N_o(\tau)^2 }{\left.2 p \N'_-\N_-\right|_0} .
\labeq{comb}
\end{eqnarray}
For reference the equal-time Lorentzian correlator then takes the
simple form
\begin{eqnarray}
G(\tau,\Omega)=\la \chi(\tau,\Omega_1)\chi(\tau,\Omega_2)\ra = \frac{\K^{3/2}}{2\pi^2}
\sum_{n=3}^\infty  G_n (\tau,\tau) \frac{n \sin n \Omega}{\sin \Omega}.
\end{eqnarray}
where $\Omega$ is the angle between $\Omega_1$ and $\Omega_2$ on the
three-sphere.


Notice that we have been able to continue $n$ by $n$ and that we have had no
convergence problems since the $\N^L$'s just oscillate; the
continuation has been much simpler than in the open case discussed
in~\cite{Gratton99}.  It has also become possible to numerically calculate
the power spectrum essentially exactly.   
We do this by making a Taylor
expansion of $\N_-$ say in the Euclidean region near the singularity at
$-X_m$, namely $ \N_- \approx 1 + \frac{(n^2-5)\K}{16\H^2} $.  We then
numerically evolve this solution from near the singularity to the
equator, and
then evolve the Lorentzian
equations until the modes are well outside the horizon and are constant.
We then compute the Lorentzian correlator for the mode.
By repeating this for each $n$ of interest, we obtain an accurate
numerical power spectrum $\P_\chi(n)$.  

We illustrate the behavior of two representative modes in the
Euclidean and Lorentzian regions in Figs.~\fig{ant2} and~\fig{ant3}.  
An example power spectrum is given in Fig.~\fig{ant4}. 
The spectrum is slightly tilted, as would be predicted for this model
using the slow-roll approximation. 
A small change in tilt is apparent on the largest scales, where the
curvature has some effect and the standard slow-roll result does not apply.


\begin{figure}[t!]
\begin{center}
\psfig{figure=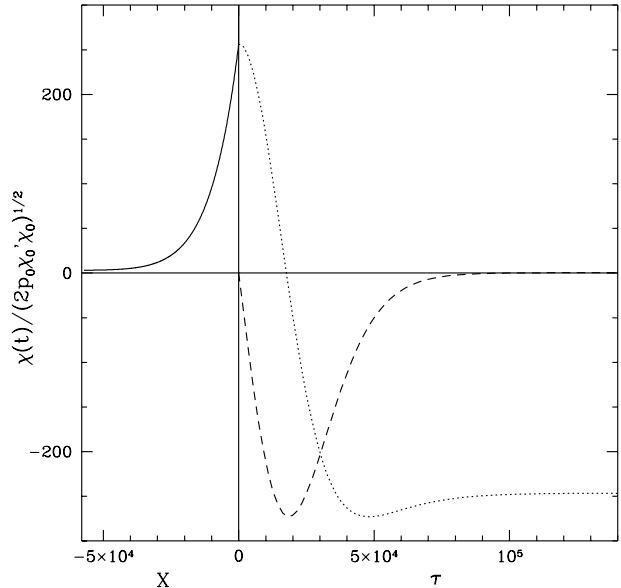,angle=0,width=8.5cm} 
\caption{The evolution of the $\nu=3$ mode, $\N(t)/\sqrt{2 p_0 \N'_0
\N_0}$. The left hand side 
shows the Euclidean conformal time evolution of the classical
solution. The RHS shows the real time evolution of the real (dotted line) and
imaginary (dashed line) parts until the modes are driven to a constant when they are
well outside the horizon.\labfig{ant2}}
\end{center}
\end{figure}

\begin{figure}[t!]
\begin{center}
\psfig{figure=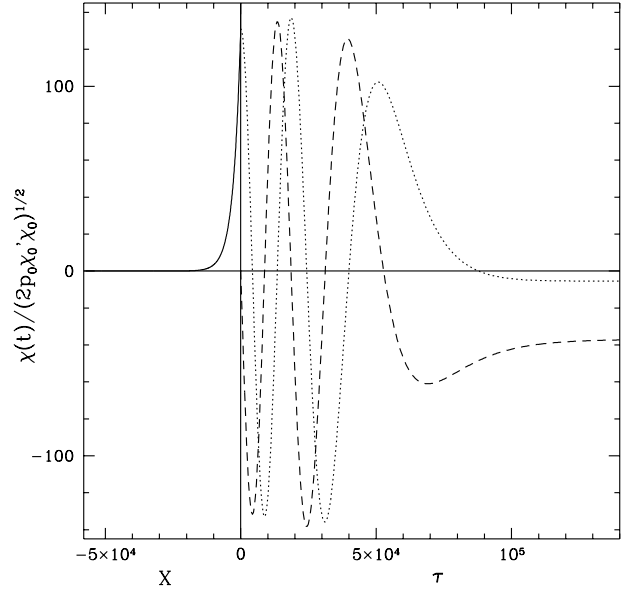,angle=0,width=8.5cm}
\caption{Evolution of the $\nu=10$ mode, axes as for Fig.~\ref{fig:ant2} \labfig{ant3}}
\end{center}
\end{figure}


\begin{figure}[t!]
\begin{center}
\epsfig{figure=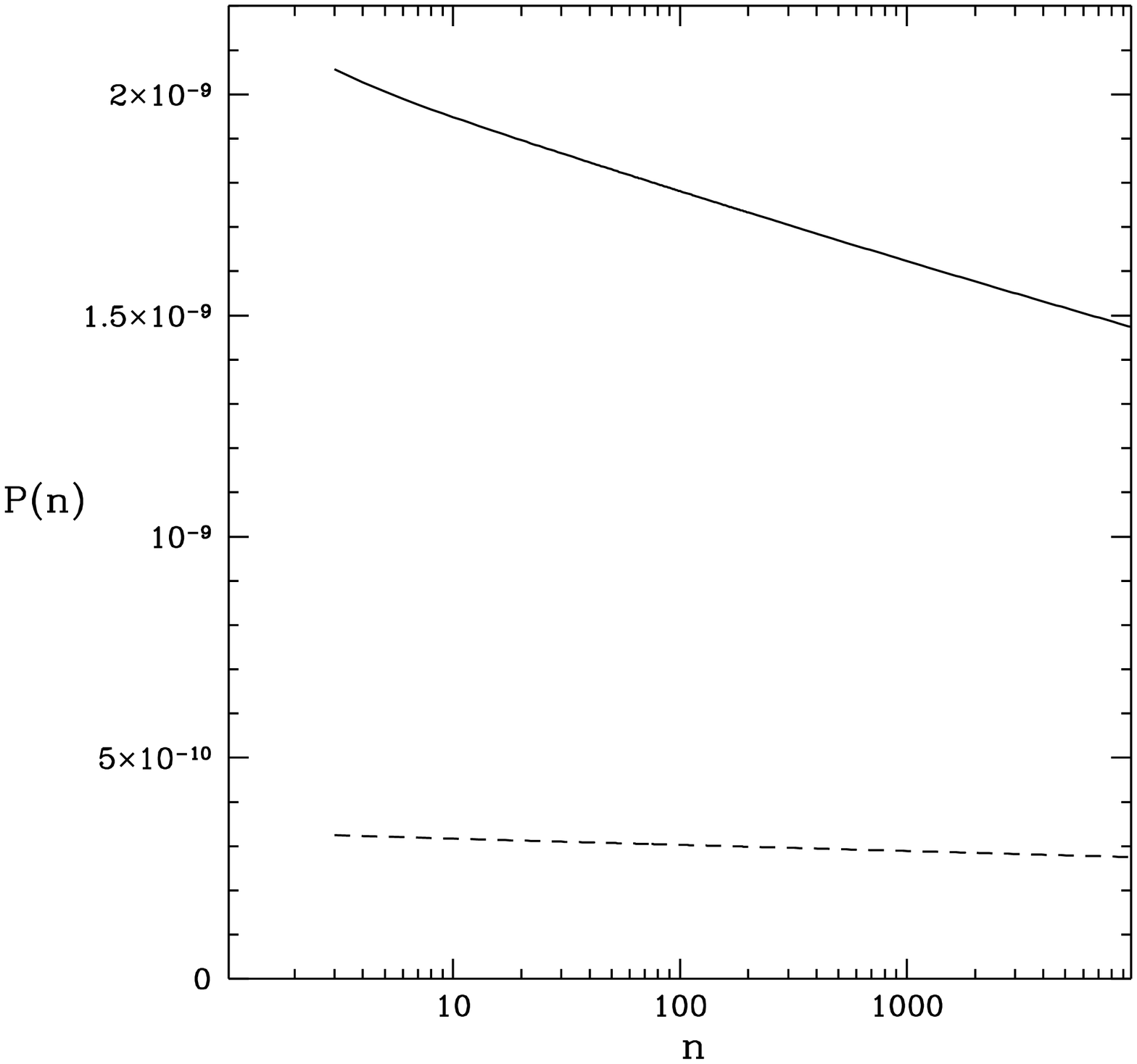,angle=-0,width=8.5cm}
\caption{The scalar power spectrum (solid line) $\P_\chi(\nu)$ and tensor spectrum
(dashed line)  $\P_h(\nu)$ for an $\half m^2\phi^2$ potential.\labfig{ant4}}
\end{center}
\end{figure}

\section{Tensors}

The derivation of the tensor power spectrum in closed universes is
very straightforward when compared to the open case, as performed in~\cite{Hertog00}.  
As with the scalars, the principle simplification is that one is able
to analytically continue mode by mode.  

The Lorentzian tensor action is
\begin{eqnarray}
S_2 = \frac{1}{8\kappa}\int d^4 x\sqrt\gamma a^2 \left[ h_{ij}'h^{ij}{}' +
(\Deltathree - 2\K)h_{ij}h^{ij}\right].
\end{eqnarray}
There are no gauge ambiguities here, and we can directly continue to
the Euclidean region, with action
\begin{eqnarray}
S_2 = \frac{1}{8\kappa}\int d^4 x\sqrt\gamma b^2 \left[ h_{ij}'h^{ij}{}' -
(\Deltathree - 2\K)h_{ij}h^{ij}\right].
\end{eqnarray}
We proceed analogously to the scalar case to calculate the Euclidean
correlator, and for
each mode we need to find the Euclidean Green function satisfying
\begin{eqnarray}
-(b^2 G_{E,n}(X,Y)')' + b^2 (n^2-1) G_{E,n}(X,Y) = \kappa \delta(X-Y),
\end{eqnarray}
which has the corresponding homogeneous equation
\begin{eqnarray}
\Phi'' + 2\H \Phi' - (n^2-1)\K\Phi = 0.
\end{eqnarray}
Here we have used the result that for tensor modes the eigenvalues of
the three-space Laplacian $\Deltathree$ are $-\K(n^2-3)$ where $n$ is integer, $n \ge 3$. The problem is
therefore very similar to the scalar case, and the same argument for
Neumann boundary conditions applies. Near a singularity the
homogeneous solution we want behaves as
\begin{eqnarray}
\Phi \approx 1 +\frac{\K(\nu^2-1)}{16\H^2},
\end{eqnarray}
and we may proceed by exact analogy with the scalar case above to
numerically calculate the symmetrized equal time tensor correlator at
the end of inflation. On super-Hubble scales the power spectrum
becomes time independent, and we define the tensor power spectrum $\P_h(\nu)$ such that
\begin{eqnarray}
\la h_{ij}(x,\tau) h^{ij}(x,\tau) \ra = \sum_\nu \frac{\nu}{\nu^2-1}\frac{\nu^2-4}{\nu^2} \P_h(\nu)
\end{eqnarray}
giving
\begin{eqnarray}
\P_h(\nu) \equiv \frac{4}{\pi^2} \K^{3/2} \nu(\nu^2-1) G_n(\tau,\tau).
\end{eqnarray}
It is shown in Fig.~\fig{ant4}.  As in the scalar case the spectrum is
close to the slow-roll prediction on all but the largest scales.

\section{CMB anisotropies}

We have shown how to calculate the  scalar and tensor power spectra at
the end of inflation. Although
we do not understand the reheating process it is well known that for adiabatic
perturbations the curvature perturbation is conserved on super-horizon
scales, as is the tensor metric perturbation.  The power spectra we have computed at the end of inflation
therefore accurately predict the curvature perturbation power spectra
for the early radiation
dominated era that we need as the starting point for the CMB
computation. We use CAMB~\cite{Lewis99}, a modified version of
CMBFAST~\cite{Seljak96}, to compute the resulting CMB anisotropies.

Since we do not know how reheating proceeds we do not know the exact
evolution of the scale factor and therefore cannot predict $\Omtot$
from the inflationary parameters to any accuracy. The observed $\Omtot$
remains a free parameter within reasonable bounds determined by the
number of e-foldings of inflation. Also, cosmological parameters such as $\Omega_{\mathrm{m}}$,
$\Omega_\Lambda$ and $H$ are not determinable without an understanding
of reheating and are effectively free input parameters into the evolution code.
Given a set of these parameters we are able to calculate the CMB
anisotropy power spectrum. An example using particular set of cosmological parameters is shown in Fig.~\fig{ant5}. 

It is important to use the correctly normalized initial power spectra
in order to get the CMB scalar/tensor ratio correct~\cite{Martin00}. In SCDM flat
models one can fix the quadrupole ratio to good accuracy analytically~\cite{Starobinsky85}. However in
general models the presence of curvature or a cosmological constant
introduces an additional integrated Sachs-Wolfe effect at low
multipoles that renders the flat result very inaccurate. In closed
models there is also a maximum wavelength cutoff which reduces the
contribution to the low multipole tensor anisotropy. To compute the
CMB power spectra correctly we computed the initial power spectra
normalized as described in this paper. The CAMB package was then updated\footnote{\url{http://camb.info}} to support
absolute computations from normalized initial power spectra so that
the scalar and tensor CMB power spectra we compute are in the correct ratio. This
improves the approximate ratio fixing scheme that was used
in~\cite{Gratton00}.

\begin{figure}[t!]
\begin{center}
\psfig{figure=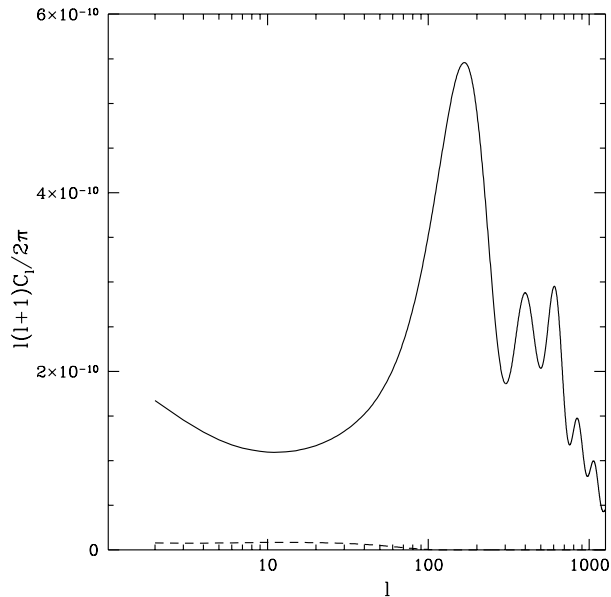,angle=0,width=8.5cm}
\caption{The total CMB temperature power spectrum (solid line) and tensor
contribution (dashed line) for an $\Omtot = 1.2$
$\Lambda$CDM model ($\Omega_\Lambda=0.8, \Omega_b=0.045, h=0.65$) with
a $\half m^2\phi^2$
potential with $m=5.4\times 10^{-6} M_{\text{pl}}$ and $\phi_0=15
M_{\text{pl}}$.\labfig{ant5}}
\end{center}
\end{figure}

\section{Conclusions}
In this paper we have shown how to calculate, from the Euclidean
path integral for Einstein gravity and a single scalar field, the scalar and
tensor power spectra for a closed universe. This was done from the saddle point approximation to the Euclidean
path integral about a constrained singular instanton.  

We have shown that the power spectra are very similar to the predictions
of flat slow-roll inflation on small scales, as expected. However we now have a method for
computing the power spectra for first order perturbations essentially exactly on all scales. Small deviations from
the slow-roll predictions are apparent only on the largest scales on
which the curvature can be significant. Using the power spectra as the
initial conditions for the radiation dominated universe we are able to 
compute predictions for CMB anisotropies in particular closed
models. Since the CMB observations are cosmic variance limited on large scales
the significance of the large scale deviations from slow-roll predictions
are limited.



\medskip
\centerline{\bf Acknowledgements}

AL thanks Anthony Challinor for very valuable discussions. AL and NT acknowledge the support
of PPARC via a  PPTC Special Program Grant. SG is supported in part by US Department of Energy
grant DE-FG02-91ER40671.


\end{document}